\documentclass[pre,final, aps]{revtex4}

\usepackage{epsfig}
\usepackage{amssymb}
\usepackage{bm}
\usepackage{graphicx}
\usepackage{amsmath}

\begin{document}

\title{Duality in  matrix lattice Boltzmann models}

\author{R. Adhikari$^{1,2}$ and S. Succi$^3$}

 \affiliation{$^1$ The Institute of Mathematical Sciences, CIT Campus, Tharamani, Chennai 600113, India. \\
 $^2$  SUPA School of Physics, The University of Edinburgh, JCMB King's Buildings, Edinburgh EH9 3JZ, United Kingdom.\\
$^3$ IAC-CNR, Viale del Policlinico, 137, I-00161 Roma, Italy.} 

\begin{abstract} 
The notion of duality between the hydrodynamic and kinetic (ghost) variables of lattice
kinetic formulations of the Boltzmann equation is introduced. It is suggested that this notion
can serve as a guideline in the design of matrix versions of the lattice Boltzmann equation
in a physically transparent and computationally efficient way.  
\end{abstract} 

\maketitle
\section{Introduction}
In the last decade, the lattice Boltzmann (LB) method  has developed into a very flexible and
effective numerical technique for the simulation of a large variety of  complex, fluid dynamical and non-equilibrium transport phenomena \cite{Succi:2001}. The LB method is based on a  stream-and-collide microscopic dynamics  of fictitious particles, which stream with a discrete set of velocities and interact according to local collision rules that drive the system towards a local equilibrium \cite{Chen:1998, Wolf-Gladrow:2000, Succi:2001}.
Mathematically, this is formulated as the lattice Boltzmann equation (LBE)
\begin{equation}\label{dbe}
\partial_t f_i + {\bf c}_i\cdot\nabla f_i = -\sum_j L_{ij}(f_j - f_j^0)
\end{equation}
where $f_{i}({\bf x},t)$ is the mean number of particles at position ${\bf x}$ and time $t$, moving along the lattice direction defined by the discrete velocity ${\bf c}_{i}, (i, j =1,...,N $). In the above, 
\begin{equation}
f_i^0 = w_i\left(\rho + {\rho{\bf v\cdot c}_i\over c_s^2} + {\rho v_{\alpha}v_{\beta}Q_{i\alpha\beta}\over 2c_s^4}\right)
\end{equation}
is a local equilibrium distribution, the discrete analogue of a Maxwellian distribution in continuum kinetic theory truncated to second order in the mean flow velocity ${\bf v}$,  the $w_i$ are a set of weights which satisfy $\sum_i w_i = 1$, and $c_s$ is the speed of sound in the LB fluid.  Greek indices denote Cartesian directions and the summation convention is implied. The low order velocity moments of the distribution function are related to the densities of mass, momentum and the deviatoric stress,
$\{\rho, \rho v_{\alpha}, S_{\alpha\beta}\} = \sum_i f_i \{1, c_{i\alpha}, Q_{i\alpha\beta}\}$
where $S_{\alpha\beta} + \rho c_s^2\delta_{\alpha\beta}= \Pi_{\alpha\beta} $ is the Eulerian momentum flux, and $Q_{i\alpha\beta}  + c_s^2\delta_{\alpha\beta}=c_{i\alpha}c_{i\beta}$.  The higher moments of the distribution are related to the densities of rapidly relaxing kinetic degrees of freedom, variously called ghost or kinetic variables. Finally, $L_{ij}$ is a scattering matrix whose eigenvalues control the relaxation of the kinetic modes to their local equilibrium values. The null eigenvalues correspond to the eigenvectors associated with the conserved mass and momentum densities, while the leading non-zero eigenvalue associated with $Q_{i\alpha\beta}$ controls the viscosity of the LB fluid.

Historically, the LBE in matrix form was derived as a Boltzmann approximation to the dynamics of lattice gas cellular automata \cite{Higuera:1989}. It was then understood that the equilibria and collision matrix could be constructed independently of the
underlying cellular automata microdynamics \cite{Higuera:1989a}, and the lattice Boltzmann approach came into being. 
The collision matrix was reduced to the simplest possible form consistent with the macroscopic hydrodynamics in \cite{Qian:1992, Chen:1992}, where the Bhatnagar-Gross-Krook (BGK) \cite{Bhatnager:1954} form of the collision term was implemented on the lattice with $L_{ij} =\tau^{-1}\delta_{ij}$.  In the lattice BGK (LBGK) model,  the fluid viscosity, which is the only transport parameter of interest, is given by 
$ \nu = c_s^2(\tau-1/2)$.   Even though it has always been clear that this simplification entails a crude approximation to the relaxation process (all modes relax at the same rate $\tau$), the LBGK equation, since its introduction,  has held the mainstream in LB applications. In a parallel development, the collision matrix version of LBE \cite{Higuera:1989a, Benzi:1992}  has been revisited, optimized, and renamed the MTR (Multiple Time Relaxation) \cite{dHumieres:1992, dHumieres:2002}, in contrast to the single relaxation time implied by the LBGK equation. A number of authors have also made a strong case  for the superiority of the MTR version over LBGK in terms of numerical accuracy and stability\cite{Lallemand:2000, McCracken:2005}. Yet, lattice BGK remains by far the most popular form of LBE to date.

The limited of popularity of the MTR approach inspite of its superiority compared to the lattice BGK method, may be due to the lack of a general guiding criterion for the spectral decomposition of the collision matrix. 
In other words, it has not been clear \emph{ a priori} how to choose the eigenvectors of the matrix $L_{ij}$ which span the kinetic space of the discrete populations $f_i$. This ambiguity arises because the conservation laws of mass and momentum fix only the hydrodynamic and transport subset of the eigenvectors, leaving the kinetic subset unspecified (see below). As a consequence, MTR models are dependent on both spatial dimension and the choice of the discrete velocity set ${\bf c}_i$,  while the LBGK model is identical across both spatial dimension and choice of velocity set. Further, the notion of orthogonality of eigenvectors in the kinetic space can itself be defined in two distinct ways: the definition followed in \cite{Benzi:1992} using a weighted inner product, and that in \cite{dHumieres:1992} using an unweighted inner product. For example, recent work on the shallow water equation \cite{Dellar:2002a}, the fluctuating lattice Boltzmann equation \cite{Adhikari:2005}, and on multireflection boundary conditions use a  set of eigenvectors which are orthogonal under the weighted inner product \cite{Chun:2007}. Clearly, it is important to understand how this non-uniqueness in the kinetic space arises and to provide a guiding principle in chosing the eigenvectors. In this work we shall propose such a guiding rule, by introducing the notion of {\it duality}  in the kinetic space of the LB.

In the next section we follow the notation in \cite{Behrend:1994} to highlight the structure of the kinetic space spanned by the eigenvectors and how a change of basis from populations to the moments reveals the dynamics of the various modes. We then introduce and illustrate the idea of duality with concrete examples. We show how a model in which all the ghost degrees of freedom are relaxed at the same rate \cite{Behrend:1994, Ladd:1994} may provide the best compromise between full MTR models where every mode has a separate relaxation time and the LBGK model. We end with a discussion on how the present work is  relevant to the algorithmic improvement of the LB method. 

\section{Spectral representation of the collision matrix}
\subsection{Eigenvectors and eigenvalues}
For a general athermal $DdQn$ LB model with $n$ velocities in $d$ space dimensions, the $n\times n$ collision matrix $L_{ij}$ has $d + 1$ null eigenvectors corresponding to the density and $d$ components of the conserved momentum,  $d(d+1)/2$ eigenvectors corresponding to the stress modes, and $n - (d + 1) - d(d + 1)/2$ eigenvectors corresponding to the ghost modes\cite{Behrend:1994, Adhikari:2005}. The choice of the null and stress eigenvectors $\{1, c_{i\alpha}, Q_{i\alpha\beta}\}$ follows directly from the physical definition of the densities associated with them. Without specifying the exact analytical expression for the remaining eigenvectors, let us label a linearly independent set of the eigenvectors of the scattering matrix by $\{A_i^a\}$, where $a = 1 \ldots n$ labels the eigenvector, and $ i = 1 \ldots n$ labels the component of the eigenvector in along the $i-th$ velocity direction. Then, we can define densities associated with the eigenvector $A_i^a$ as moments of the populations by
\begin{equation}
\psi^a({\bf x}, t) = \sum_i f_i({\bf x}, t) A_i^a
\end{equation}
For $A^a_i =\{1, c_{i\alpha}, Q_{i\alpha\beta}\}$ the densities are the mass, momentum and stress. The ghost eigenvectors are higher polynomials of the discrete velocities \cite{Benzi:1992} . The discreteness of the kinetic space implies that, unlike in the continuum, only a finite number of polynomials can be linearly independent, being equal to the number of discrete velocities. For a model with $n$ discrete velocities, the choice of the $n$ linearly independent polynomials is thus not unique, but defined only upto a similarity transformation. Thus, the reason for the non-uniqueness in the spectral decomposition can be traced to the discreteness of the velocity space itself . Independent of the precise choice, the distribution function itself can be expanded in a linearly independent set eigenvectors which are polynomials of the discrete velocities
\begin{equation}\label{fexpansion}
f_i({\bf x}, t) = w_i\sum_a \psi^a({\bf x}, t){ A_i^a\over N^a}
\end{equation}
Consistency between the above two equations implies that the set of polynomials $A_i^a$ are both orthogonal and complete,
\begin{eqnarray}
&\sum_i & w_i A_i^a A_i^b = N^a\delta^{ab},\\
&\sum_a & A_i^a A_j^a/N^a = \delta_{ij}.
\end{eqnarray}
Crucially, with the definitions above \cite{Adhikari:2005}, the eigenvectors $A_i^a$ form an orthogonal set under an inner product $(A^a, A^b) =\sum_i w_i A_i^a A_i^b $.  This inner product is identical to that introduced by Benzi \emph{et al} \cite{Benzi:1992}, but distinct from the unweighted inner product $(A^a, A^b) =\sum_i A_i^a A_i^b $ used by d'Humieres and co-workers\cite{dHumieres:1992, dHumieres:2002}. The advantages of the present choice are discussed below. As indicated before, a useful categorisation of the polynomials consists of the $d+1$ polynomials $\{1, c_{i\alpha}\}$ corresponding to the mass and momentum, the $d(d+1)/2$ quadratic polynomials $Q_{i\alpha\beta}$ corresponding to the stress, and the remaining $n - (d+1) - d(d+1)/2$ cubic and higher order polynomials corresponding to the ghost variables. Correspondingly, the distribution function can be separated into contributions from the hydrodynamic, transport, and ghost moments
\begin{equation}
f_i = f_i^H + f_i^T + f_i^G.
\end{equation}
This motivates the introduction of projection operators  \cite{Behrend:1994} which project the distribution function onto the hydrodynamic, transport, and ghost subspaces,
\begin{eqnarray}
\sum_j P^H_{ij} f_j &=& f_i^H = w_i\left(\rho + {\rho{\bf v}\cdot{\bf c}_i\over c_s^2} \right)\\
\sum_j P^T_{ij} f_j &=& f_i^T = w_i{S_{\alpha\beta}Q_{i\alpha\beta}\over 2 c_s^4}\\
\sum_j P^G_{ij} f_j &=& f_i^G = w_i\sum_{a\in G} \psi^a A_i^a/N^a\\
\end{eqnarray}
The explicit form of the projection operators are
\begin{eqnarray}
P^H_{ij} &=& w_i(1 + {\bf c}_i\cdot{\bf c}_j/c_s^2)\\
P^T_{ij} &=& w_iQ_{i\alpha\beta}Q_{j\alpha\beta}/2c_s^4 \\
P^G_{ij} &=&\sum_{a\in G} w_i A_i^a A_j^a/N^a
\end{eqnarray}
The discrete Maxwellian is a nonlinear (quadratic) function of the distribution function, and thus Eq.\ref{dbe} is a only apparently linear, the nonlinearity being concealed in $f_i^0$.  An useful linearisation of the LB equation consists of neglecting the quadratic term in the discrete Maxwellian to yield a local equilibrium $h_i^0$ which is linear in the mean velocity,
\begin{equation}
h_i^{0} = w_i\left(\rho + {\rho{\bf v\cdot c}_i\over 2 c_s^2}\right) = \sum_j P_{ij}^H f_j
\end{equation}
In the linearised approximation for the equilibrium distribution, we have $f_i^0 = h_i^0 = (P^H f)_i$ and so the 
linearised LBE can now be written as,
\begin{equation}
\partial_t f_i + {\bf c}_i\cdot\nabla f_i = -\sum_j L_{ij}[f_j - (P^H f)_j]= -\sum_j L_{ij}^Rf_j
\end{equation}
where $L_{ij}^R = \sum_k L_{ik}(1 - P^H)_{kj}$ is a right-projected collision matrix. In this form, it is clear that $L_{ij}^R$ by construction has eigenvectors of mass and momentum with zero eigenvalues.  The form of the matrix,  by itself, places no constraint on the eigenvalues of transport and ghost sectors. However, the requirments of an extended range of hydrodynamic behaviour, stability and isotropy motivate an optimal construction of $L_{ij}^R$. As explained in the Introduction, the simplest possible model consists of a diagonal collision matrix $L_{ij} = \delta_{ij}/\tau$ which implies that all the non-conserved modes relax at the same rate $1/\tau$. This
is the very popular LBGK approximation used in the literature. In the hydrodynamic regime, a scale separation exists between the relaxation of the conserved and non-conserved variables: the mass and momentum densities relax slowly, the stress and ghost variables relax rapidly.  One variant of a model used by Ladd \cite{Ladd:1994} uses adjustable relaxation times for the stress modes, and identical unit relaxation times for the ghost modes, i.e. the ghosts are `projected' out. One advantage of this approach is that the precise form of the ghost modes, which in general differ both in number and in form between LB models, need not be known. A generalisation of this model, with two relaxation times \cite{Behrend:1994}  reads,
 \begin{equation}
L_{ij}^R = \lambda P_{ij}^T + \sigma P_{ij}^G = \sigma(1 - P_{ij}^H) + (\lambda -\sigma)P_{ij}^T
\end{equation}
where the last follows from the completeness relations $P^H + P^T + P^G = 1$. Since the precise form of the ghost projection operator, and hence the ghost eigenvectors is never needed in this formulation, it is clear that the linearised dynamics in this two-relaxation time model cannot depend on the precise choice of the ghost mode eigenvectors. The only way this model may be optimised is to tune the relaxation rate of the ghost modes in comparision to the stress modes. However, a model which allows separate relaxation times for each individual ghost mode has a greater flexibility and may be optimised to yield the best range of hydrodynamic behaviour \cite{dHumieres:2002}. It needs careful analysis to see if the gain is enough to justify the loss of simplicity and generality that one obtains from the two relaxation model. Hydrodynamic behaviour is obtained when there are two propagating modes with a dispersion relation 
$\omega = c_s k + i\nu_L k^2$, $\nu_L=\nu + 3/2 \nu_{bulk}$ being the longitudinal viscosity, and $d-1$ diffusive modes with a dispersion relation $\omega = i\nu k^2$. Both the speed of sound and the viscosities are assumed to be constant. 

\subsection{Linear mode structure}
The hydrodynamic behaviour of the linearised LBE is most conveniently analysed in the absence of boundaries when a Fourier mode decomposition is possible \cite{Das:1993, Behrend:1994, Lallemand:2000, Dellar:2002a}. It is important to note that the departure from hydrodynamic behaviour can arise from two distinct sources. The first is the choice of eigenvectors and relaxation times of the discrete velocity (but space and time continuous) LBE. This is the category of error arising from discretisation in velocity space. The second is that arising from the numerical integration of the LBE. This is the category of error arising from discretisation in space and time. The physical and numerical behaviour of the fully discretised LBE dynamics is a combination of both these sources of error. The present work, focussing as it does only on the kinetic space, has direct implications for errors arising out of discretisation of velocity space. The errors arising out of discretisation of space and time are relatively well understood from the numerical analysis of the hyperbolic differential equations. In particular, it is known that an Euler integration step of size $\Delta t$ produces numerical diffusion, and thereby renormalises the viscosity to  $\nu = c_s^2(\tau - \Delta t/2)$ \cite{Chen:1998}. 

To derive the dispersion relation we  Fourier transform the linearised LBE to get
\begin{equation}\label{linlbe}
\partial_t f_i + i{\bf k}\cdot{\bf c}_i f_i = -\sum_j L_{ij}^Rf_j
\end{equation}
At ${\bf k=0}$, the eigenmodes of the dynamics are the same as the eigenmodes of $L^R$. However, away from ${\bf k=0}$, neither the eigenmodes nor the eigenvalues are identical. For small $k$, an analytical expression for the eigenvalues may be
obtained perturbatively \cite{Behrend:1994} . For arbitrary $k$, a numerical solution is necessary.  The dispersion relation is obtained by a Fourier transform in time,
\begin{equation}
-i\omega({\bf k})f_i = -\sum_j [i{\bf k}\cdot{\bf c}_i \delta_{ij} + L_{ij}^R]f_j
\end{equation} 
Thus we need to obtain the eigenvectors and eigenvalues of the matrix
\begin{equation}
M_{ij} = i{\bf k}\cdot{\bf c}_i \delta_{ij} + L_{ij}^R.
\end{equation}
The dynamics in Eq.\ref{linlbe} can equally well be written in terms of the densities using Eq.\ref{fexpansion} as
\begin{equation}
\partial_t \psi^a = -\sum_b[\Gamma^{ab} + \lambda^a\delta^{ab}]\psi^b
\end{equation}
where matrix coupling the different modes is
\begin{equation}
N^a\Gamma^{ab} = i{\bf k}\cdot\sum_i w_i A^a_i A^b_i {\bf c}_i
\end{equation}
It is worth noting that the linearised LB dynamics can be written in either of the forms
\begin{eqnarray}
\partial_t f_i &=& -\sum_j (\mathcal{A + C})_{ij}f_j\\
\partial_t \psi^a &=& -\sum_b (\mathcal{A + C})^{ab}\psi^b
\end{eqnarray}
The dynamical equation in the $f_i$ basis diagonalises the advection operator $\mathcal{A}_{ij} = i{\bf k\cdot c}_i \delta_{ij}$ , while the dynamical equation in the $\psi^a$ basis diagonalises the collison operator $\mathcal{C}_{ij} = L_{ij}^R$. The eigenvectors of the dynamics are a combination of the $f_i$ and the $\psi^a$.  The dispersion relation equation can be conveniently non-dimensionalised by measuring time in units  of the inverse of the relaxation rate for the stress modes $\tau = \lambda^{-1}$, and distance in units of $c_s\tau$. The non-dimensionalised dispersion equation then takes the form
\begin{equation}
-i\Omega({\bf q})f_i = -\sum_j [i{\bf q}\cdot{{\bf c}_i\over c_s} \delta_{ij} + {\cal L}_{ij}^R]f_j
\end{equation} 
where $\Omega = \omega\tau$ is a non-dimensionalised frequency and ${\bf q} = {\bf k}c_s\tau$ is a non-dimensionalised wavevector. It should be noted that the non-dimensionalised collision matrix ${\cal L}^R = L^R/\tau$ now depends only on the ratio $\sigma/\lambda$ of the relaxation rates of the ghost and stress eigenvectors. We shall use this non-dimensionalised form of the dispersion relation to obtain the numerical eigenspectrum of  one  of the LBE models presented below.

\section{Duality in lattice kinetic theory}

The symmetry principle of duality, which relates two different mathematical representations of the same physical theory, is a powerful tool in many areas of physical science. Duality is often use to map strongly interacting degrees of freedom to weakly interacting ones, thus facilitating an approximate, and often,  even an exact solution of the problem. A celebrated example is the solution of Kramers and Wannier for the critical temperature of the Ising model \cite{Kramers:1941}.  To the best of our knowledge, dual symmetries do not appear to have played any major role in kinetic theory. In the context of the lattice Boltzmann schemes, we introduce duality not as an exact symmetry, but as a requirement on the structure of the kinetic space of the theory. Specifically, we require that the structure of the ghost subspace should mirror that of the hydrodynamic subspace, and consist of scalar densities and associated vector currents. In our notation, a LB kinetic space is exactly dual if each ghost field corresponds to a hydrodynamic field and a suitable transformation converts the ghost degrees of freedom into hydrodynamic degrees of freedom. If this exact correspondence is broken, but the ghost subspace still consists of sets of scalar densities and vector currents we say that the kinetic space is quasi-dual. The scalar densities and vector currents are taken to be even and odd functions of the discrete velocities respectively. Thus introduced, duality is a \emph{normative} principle on the structure of the kinetic space of the LBE.  The duality principle, as we show with several examples below, allows us to choose the eigenvectors of the collision matrix in a way which is both transparent and unique. 

\subsection{Two dimensions}
Let us first consider the standard $D2Q9$ model with the usual set of velocities connecting the four nearest neighbours and the four next-nearest neighbours of the square lattice. Thus there are four velocities with unit modulus, another four with modulus two, which together with the zero velocity give the nine dynamical populations of the $D2Q9$ model. The kinetic space is spanned by eigenvectors corresponding to the mass and momentum, $\{A^{0}_i, A^{1}_i, A^{2}_i\}  = \{1_i, c_{ix}, c_{iy}\}$. The next three natural eigenvectors associated with stress tensor are $\{A^{3}_i, A^4_i, A^5_i\}$ = $\{Q_{ixx}, Q_{ixy}, Q_{iyy}\} \equiv \{ c_{ix}^2-c_s^2, c_{ix}c_{iy}, c_{iy}^2-c_s^2\}$.  All of these are recognized as discrete velocity analogues of tensor Hermite polynomials \cite{He:1997e} .  Without any physical considerations to guide us, the choice of three higher-order eigenvectors, associated with the ghost modes remains open. An obvious choice is the next series of tensor Hermite polynomials, that is $Q_{ixx} c_{ix}$, $Q_{ixx} c_{iy}$, $Q_{iyy} c_{ix}$, $Q_{iyy} c_{iy}$. It is immediately seen that due the identity $c_{ix}^3 = c_{ix}$, holding for the $D2Q9$ lattice, only two of these are linearly independent. This lack of linear independence,  as we mentioned earlier, is due to the discrete nature of the velocities, giving identities like $c_{ix}^3 = c_{ix}$, which are absent in the continuum.  To complete the kinetic space, one more eigenvector is required. It is immediately checked that, as a consequence of the $D2Q9$ identity $c_{ia}^4 = c_{ia}^2$, $a=x,y$, out the five Hermite polynomials of order 4, only one is linearly independent, which we chose as $Q_{ixx} Q_{iyy}$. This then completes the construction of the remaining three ghost eigenvectors.

 In a very illuminating paper, Dellar \cite{Dellar:2002a} proposes a different decomposition, based on the notion of ghost densities introduced in \cite{Benzi:1992}. The first ghost eigenvector is of the form
\begin{equation}
G^{0}_i \equiv g_i=(1,-2,-2,-2,-2,4,4,4,4) 
\end{equation}
and the remaining two are simply the corresponding `currents', that is
\begin{equation}
G^{1}_i \equiv g_i c_{ix}, 
G^{2}_i \equiv g_i c_{iy}
\end{equation}
The physical meaning of this choice is best highlighted by expressing $g_i$ in analytical
form, that is
\begin{equation}
g_i= { c_i^4\over 2c_s^4} -{5 c_i^2\over  2 c_s^2} + 1_i
\end{equation}
where $c_i^2 = c_{ix}^2 + c_{iy}^2$.
It is easily checked that the basis $A^{0} \dots A^{5}, A^6 = G^{0}, A^7 = G^{1}, A^8 = G^{2}$ is orthogonal under the weighted scalar product $(A^a,A^b) = \sum_i w_i A_i^a A_i^b$, where $w_0=4/9$, $w_{1-4}=1/9$ and $w_{5-8}=1/36$ are the usual $D2Q9$ weights. 
It is also to be noted that, owing to the D2Q9 identities, the ghost eigenbasis can also be written as
\begin{widetext}
\begin{eqnarray}
G_i^0=  c_{ix}^2 c_{iy}^2 - (3/2) (c_{ix}^2+c_{iy}^2) + 1\\
G_i^1 = c_{ix} c_{iy}^2 - (3/2) (c_{ix}+c_{iy}^2 c_{ix} ) + c_{ix}=-(1/2) c_{ix}(1+c_{iy}^2)\\
G_i^2 = c_{iy} c_{ix}^2 - (3/2) (c_{iy}+c_{ix}^2 c_{iy} ) + c_{iy}=-(1/2) c_{iy}(1+c_{ix}^2)
\end{eqnarray}
\end{widetext}
Surprisingly, then, $G^0_i = g_i$ is a fourth order lattice Hermite polynomial, while $G_i^1 = g_i c_{ix}$ and $G_i^2 = g_i c_{iy}$ instead of being fifth order lattice Hermite polynomials turn out to be third order lattice Hermite polynomials. This fulfils exactly the duality principle introduce above: the kinetic space is decomposed into a set of eigenvectors corresponding to conserved, transport and ghost moments; the ghost degrees of freedom correspond to an even scalar density and two odd vector currents and are in one-to-one correspondence with the hydrodynamic degrees of freedom; and as we show below, the ghost and hydrodynamic degrees of freedom are related by a suitable transformation.

The duality in the decomposition is beautifully illustrated by the diamond structure of the $D2Q9$ eigenvectors shown in Table 1. The density and the two momenta are matched by a ghost density and two ghost currents. The dynamical behaviour of these degrees of freedom are of course quite different, as is revealed by displaying the LBE dynamics in the basis of moments. The kinetic moments associated with the present choice of eigenvectors is
\begin{equation}
\{\rho, \rho v_{\alpha}, S_{\alpha\beta}, \rho^{\prime},  j^{\prime}_{\alpha}\} = \sum_i f_i \{1, c_{i\alpha}, Q_{i\alpha\beta}, g_i, g_i c_{i\alpha}\}
\end{equation} 
The primed quantities correspond to ghost density and its currents. The decompostion of  $f_i$ as the sum of a hydrodynamic, transport  and ghost components is
\begin{eqnarray}
f^H_i &=& w_i( \rho +{{\bf j\cdot c}\over c_s^2})\\
f^T_i &=& w_i( {S_{\alpha\beta} Q_{i\alpha\beta} \over 2c_s^4})\\
f^G_i &=& {1\over 4} w_i g_i({\rho^{\prime}} + {{\bf  j}^{\prime}\cdot{\bf c}\over c_s^2})
\end{eqnarray}
From the above expressions it is clear that, to within a scale factor, the ghost sector is transformed into the conserved sector under the duality transformation
$1_i \leftrightarrow g_i$.

To physically interpret the above decomposition, we first note that the combination $w_i^\prime = w_i g_i$ may be interpreted as the weight associated with the ghost degrees of freedom. Then, the weights of the hydrodynamic modes sum to unity $\sum w_i = 1$, while the weights of the ghost modes sum to zero $\sum_i w_i^\prime = 0$. This last result combined with the fact that ghost density is even in the velocities $g_i = g_{i^{\star}}$, where ${\bf c}_{i^{\star}} = -{\bf c}_{i}$, indicates that the ghosts correspond to oscillatory eigenvectors familiar in quantum and statistical mechanics, where they represent excitations above the ground state or above equilibrium. The ghost degrees of freedom are thus non-equilibrium excitations carried by even, oscillatory eigenvectors. The even and odd character of eigenvectors can be exploited to classify the entire set of moments into two categories: even moments representing densities, and odd moments representing currents. Odd moments, representing currents, vanish at global equilibrium by symmetry. The even moments are not constrained to vanish by symmetry arguments. However, since the kinetic modes have no projection onto the global equilibrium distribution function $f_i^0 = w_i\rho$, they can be conveniently chosen to vanish at equilibrium. This is one of the principal advantages of using a set of eigenvectors which are orthogonal under the weighted inner product. 

By interpreting $w_i$ as `masses' of the hydrodynamic modes, the  $w^{\prime}_i$ can be identified with `masses' of ghost modes. By construction, since they sum up to zero, some of these masses ought to be negative. For instance, the ghost density can be rewritten as an alternating sum of the populations associated with the three energy levels $c_j^2 =0,1,2$, that is $\rho^{\prime} = f_0 - 2(f_1 + f_2 + f_3 + f_4) + 4(f_5 + f_6 + f_7 + f_8) = \rho_0 -2 \rho_1 + 4 \rho_2$, where $j=0,1,2$ refer to the $j$-th energy level. Being the sum of populations, each of the partial densities is strictly non-negative at all times, but the combination of alternating coefficients is a signed quantity, $\rho^{\prime}$, which is zero only at equilibrium. The duality is made even more apparent by defining the reduced distribution function $\phi_i \equiv f_i/w_i$, thus writing $\rho = \sum_i w_i \phi_i$ and  $\rho^{\prime} = \sum_i w_i^{\prime} \phi_i$. Since $w_i$ is the lattice analogue of global equilibrium distribution, $w_i^{\prime}$ may also be interpreted as a measure of the global departure from equilibrium.

The ghost currents are a measure of the skewness of the kinetic distribution function, which is non zero only out of equilibrium. Being based on this equilibrium $\leftrightarrow $ non-equilibrium duality, the ghost decomposition shows that the higher-order excitations, keeping the system away from equilibrium, can be structured exactly like their hydrodynamic counterparts. It should be appreciated the duality is structural and not dynamical: it is broken at various levels, starting with the prefactors defining the ghost density and current, because the norm of the dual hydrodynamic versus ghost eigenvectors is not the same. In particular, this implies that the ghost kinetic tensor is not isotropic, as one can easily check by a direct calculation: $P_{xx}^{\prime} = 4 P_{xx}$ and $P_{xy}^{\prime} = - 4 P_{xy}$. This is not surprising, since equilibrium and non-equilibrium are {\it not} physically equivalent.

However, a dynamical transformation in time,  $\lambda \leftrightarrow 1/\lambda$ turns perfectly conserved modes (infinite-lifetime) into perfectly {\it non}-conserved ones (zero-lifetime). What this means is that the distinction between equilibrium and non-equilibrium modes is not dictated by the structure of the kinetic space, but only by the actual  values of the lifetimes of the excitations supported by this equation. In this respect, we expect a signature of this dynamical duality in the form of a mirror symmetry ${\lambda/\sigma}\leftrightarrow {\sigma/\lambda}$ in the  dispersion relation for the two-relaxation time model introduced earlier.  Numerical dispersion relations presented in the next section do show evidence of such a symmetry. 

\begin{table}
\begin{tabular}{ccccccccc}
 &  &  &  & 1 &  &  &  & C \\ 
 &  &  & x &  & y &  &  & C \\ 
\hline
 &  & xx &  & xy &  & yy &  & T \\ 
\hline
 &  &  & xyx &  & xyy &  &  & G \\ 
 &  &  &  & xxyy &  &  &  & G \\ 
\end{tabular}
\caption{The arrangement of the eigenvectors of the $D2Q9$ lattice in a diamond structure. There is an exact duality about the transport sector (T) with the conserved hydrodynamic (C) and ghost (G) sectors, transforming into each other under the interchange of weights (see text).}
\end{table}

The structural duality of the kinetic space is broken dynamically by the different eigenvalues  assigned to the hydrodynamic and ghost modes. The hydrodynamic sector sustains itself even in the absence of ghost (standard macroscopic hydrodynamics), whereas the ghosts, because of finite lifetimes assigned to them,  do not survive without a forcing from the hydrodynamic modes.  Indeed, in the absence of the hydrodynamic feedback on $P^{\prime}_{\alpha \beta}$, the ghost sector would rapidly estinguish, because neither its density nor its current are conserved in time. Of course, such dynamical asymmetry can always be removed by choosing the ghost eigenvalues equal to be zero. In fact, there is even some evidence that long-lived ghosts may prove beneficial to the numerical stability of short-scale hydrodynamics in fluid turbulence \cite{Sbragaglia:2006a}. This is a prescription to keep the Boltzmann distribution away from equilibrium for an indefinitely long time, leading to anomalous relaxation. This prescription clearly violates the normal ordering between slow, hydrodymic and fast, kinetic modes, as it corresponds to enforcing additional conservation laws with  no counterpart in the real molecular world. Hence, such a procedure can only be justified as an effective interaction between collective degrees of freedom, as for example in lattice kinetic equations for turbulent flows.

\subsection{Higher order lattices in two dimensions}
In two dimensions, lattices with more than nine velocities are used in thermal LBE and in applications in microfluidics and multiphase flow. Duality can be used to generate an optimal kinetic space for these higher order lattices as well. Let us denote by $D(s)$ the lattice 
corresponding to a hierarchical tree of eigenvectors, with $2s + 1$ levels and symmetric about the $s+1$th level.(see Table 2 and 3). 
Clearly, this hierarchy contains $(s + 1)^2$ independent moments. With this definition,  $D(0)$ corresponds to the lattice with a single zero-speed rest particle, $D(1)$ to the $D2Q4$ lattice, and $D(2)$ with the $D2Q9$ lattice. $D(3)$, which represents a higher order lattice in the present terminology, consists of $16$ velocities with four velocities each of modulus $1, 2, 4$ and $8$. For this rather complicated lattice, the duality prescription proceeds by first chosing the usual eigenvectors corresponding to the conserved and transport sectors.  Proceeding to the $3rd$ level, the eigenvectors of the type $c_{i\alpha}Q_{i\beta\gamma}$  (and permutations) now turn out to be linearly independent. The remaining moments are constructed in a top-down fashion, beginning with a sixth-order scalar density $\rho_{6i} = A c_i^6 - B c_i^4 + C c_i^2 - 1_i$ from which two currents $\rho_{6i}c_{i\alpha}$ and three tensorial densities $\rho_{6i}Q_{ixx}, \rho_{6i}Q_{ixy}, \rho_{6i}Q_{iyy}$ can be used to complete the hierarchy. The expansion coefficients $A, B, C$  for the scalar density can be computed by requiring orthogonality to the lower eigenvectors. 

\begin{table}
\begin{tabular}{ccccccccc}
&  &  & 1 &  &  &  & C &  lattice \\ 
&  & x &  & y &  &  & C & Hermite\\ 
& xx &  & xy &  & yy &  & T & expansion  \\ 
\hline
xxx &  & xxy &  & xyy &  & yyy & T & \\ 
\hline
& $\rho_6$xx &  & $\rho_6$xy &  & $\rho_6$yy &  & G &  \\ 
&  & $\rho_6$x &  & $\rho_6$y &  &  & G & duality \\ 
&  &  & $\rho_6$ &  &  &  & G &  \\ 
\end{tabular}
\caption{The hierarchical tree of moments for the 16 speed dual lattice $D(3)$ defined in the text. The first four levels are constructed using
the usual lattice Hermite polynomials. The remaining three levels are completed using the duality prescription starting with an even, sixth-order scalar density}
\end{table}

The next member of the hierarchy is $D(4)$, corresponding to the lattice of $25$ speeds which has recently been shown to have $8$th order isotropy in its spatial behaviour. As with $D(3)$, the eigenvectors upto the $3rd$ level are the conserved, transport, and tensorial Hermite polynomials  $c_{i\alpha}Q_{i\beta\gamma}$ and permutations. Again, on the $25$ velocity lattice, the permutations give rise to independent eigenvectors. The remaining eigenvectors constructed bottom-up starting from an even, scalar eighth-order density $\rho_{8i} = A c_i^8 - B c_i^6 + C c_i^4 - Dc_i^2 +1_i$, from which two currents $\rho_{8i}c_{ix}$, s $\rho_{8i}c_{iy}$, three tensorial densities $\rho_{8i}Q_{ixx}$, $\rho_{8i}Q_{ixy}$, $\rho_{8i}Q_{iy}$, and four tensorial currents $\rho_{8i}t_{ixxxx}$, $\rho_{8i}t_{ixxyy}$, $\rho_{8i}t_{ixyyy}$, $\rho_{8i}t_{iyyyy}$ can derived, thus completing the list of $25$ independent eigenvectors.

\begin{table*}
\begin{tabular}{cccccccccccc}
 &  &  &  &  & 1 &  &  &  &  & C & lattice \\ 
 &  &  &  & x &  & y &  &  &  & C & Hermite \\ 
 &  &  & xx &  & xy &  & yy &  &  & T & expansion \\ 
 &  & xxx &  & xxy &  & xyy &  & yyy &  & T &  \\ 
\hline
 & xxxx &  & xxxy &  & xxyy &  & xyyy &  & yyyy & T &  \\ 
\hline
 &  & $\rho_8$xxx &  & $\rho_8$xxy &  & $\rho_8$xyy &  & $\rho_8$yyy &  & G &  \\ 
 &  &  & $\rho_8$xx &  & $\rho_8$xy &  & $\rho_8$yy &  &  & G & duality \\ 
 &  &  &  & $\rho_8$x &  & $\rho_8$y &  &  &  & G &  \\ 
 &  &  &  &  & $\rho_8$ &  &  &  &  & G &  \\ 
\end{tabular}
\caption{The hierarchical tree of moments for the $25$ speed dual lattice $D(4)$ defined in the text. The first four levels are constructed using
the usual lattice Hermite polynomials. The remaining three levels are completed using the duality prescription starting with an even, eighth-order scalar density}  
\end{table*}

Both of the above examples show that the duality prescription offers a transparent method of choosing and ordering the set of eigenvectors in lattices which are more complicated than the most commonly used $D2Q9$ lattice in two dimensions. It has recently been shown that the $16$ and $25$ speed lattice with a proper choice of weights, provide $6$th and $8$th order isotropy respectively. 
\cite{Sbragaglia:2007,Shan:2006}. Since the choice of weights is intimately related to both weighted inner product and the duality prescription, it is possible that there is fundamental link between isotropy and the duality prescription. 

\subsection{Three dimensions}

In three dimensions, the model which is potentially exactly dual is the $D3Q14$ model, which has the usual $4$ hydrodynamic degrees of freedom, $6$ transport degrees of freedom, leaving $4$ ghost degrees of freedom to match the hydrodynamic ones. However, since the $D3Q14$ model is not used in practice, we pass on instead to the analysis of the most common $D3Q19$ model. Here, of course, the kinetic space can only be quasi-dual, since there are $9$ ghost degrees of freedom. To choose them according to the duality prescription, several possibilities can be explored. With one ghost density quartic in the velocity, and three associated currents which are quintic, we obtain $4$ ghost eigenvectors,  leaving $5$ free. This allows three independent components of the ghost momentum-flux tensor ($xy,xz,yz)$, plus another two, which must necessarily come from a higher Hermite level. This seems to be a rather obscure and unpromising avenue.  A better possibility is to select {\it two} ghost densities along with their currents, leaving the third one `naked', \emph{i.e.} without independent degrees of freedom for the current. Retaining {\it three} quartic ghost densities with  their respective currents is unviable, for it gives a total of twelve eigenvectors, three too many.  From these considerations, it appears that the $2$ (dressed) plus $1$ (naked) density representation comes closest to fulfillingl the duality program. It is interesting to note that, apart from the third, naked, density, this is precisely the early decomposition adopted in \cite{Benzi:1992}, based on the 3d projection of 4d face-centered hypercube (24 speeds in $d=4$, 18 in $d=3$). The explicit form of the chosen eigenvectors is given in the Appendix.

\section{Numerical results}
We now present a numerical calculation of the dispersion relation of the two-relaxation time lattice Boltzmann model with the duality-prescribed choice of eigenvectors. The dispersion relation is obtained by numerically computing the eigenvalues and eigenvectors of the matrix $M_{ij}$. 
As explained previously, with a suitable rescaling, the only parameter in the problem is $\sigma/\lambda$, the ratio
of the relaxation rates of the kinetic and stress degrees of freedom. For $\sigma = \lambda$ the collision term
reduces to the LBGK diagonal collision operator. The imaginary parts of the eigenvalues for the D$2$Q$9$ model
are shown in Fig.1 , clearly showing the presence of hydrodynamic modes (relaxation rates vanish
as wavenumber goes to zero) as well as non-hydrodynamic modes 
(relaxation rates remain finite as wavenumber goes to zero). 
The scale separation between the relaxation rates of the hydrodynamic and non-hydrodynamic modes
becomes progressively smaller with increasing wavenumber, and there is considerable overlap at 
around $q = \pi$. This is fairly plausible, since $q = \pi$ is the value
at which the wavelength becomes comparable with the mean free path $c_s \tau$, so that the distinction
between hydrodynamics and kinetic modes fades away.  
This lack of scale separation is responsible for the poor range of hydrodynamic behaviour of the LBGK models,
a fact that was correctly noted earlier \cite{Lallemand:2000}. 
With $\sigma = {1\over 2}\lambda$, the overlap between the hydrodynamic and non-hydrodynamic
modes in Fig.2 is even greater, indicating a further reduced range of hydrodynamic behavior, compared to the LBGK models. On
the other hand, for $\sigma = 2\lambda$, we see in Fig.3 a clean separation between 
the hydrodynamic and kinetic degrees of
freedom, and it is in this range of parameters that we expect the best hydrodynamic 
behavior of the two-relaxation time LB model. 
An interesting qualitative feature that emerges from comparing 
Fig.2 and Fig.3, is that the eigen-frequencies 
are almost `dual' to each other, in the sense that the dispersion curves are 
approximately identical after a reflection about the ordinate and a rescaling by $\lambda\over\sigma$.  The structural
duality reflects itself in the dynamical behaviour if the relaxation times are chosen appropriately. 

\begin{figure}
\includegraphics[height = 6cm]{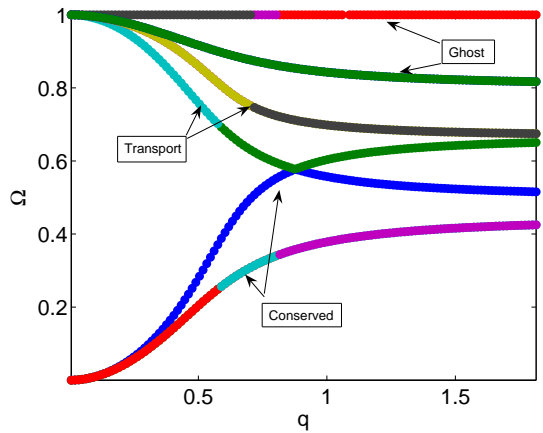}
\caption{Dispersion relation of the $D2Q9$ matrix LBE for $\tau = \sigma = 1$. This is identical to the BGK model.
Note the large overlap of the hydrodynamic (lower curves) and non-hydrodynamic (upper curves) modes.}
\end{figure}

\begin{figure}
\includegraphics[height = 6cm]{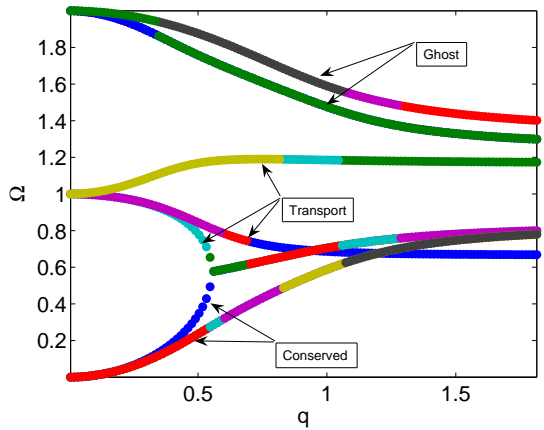}
\caption{Dispersion relation of the $D2Q9$ matrix LBE for $\tau = 1$, $\sigma = 2$. The ghost modes are forcibly
made to relax slower than the stress modes, leading to poor hydrodynamic behaviour.}
\end{figure}

\begin{figure}
\includegraphics[height = 6cm]{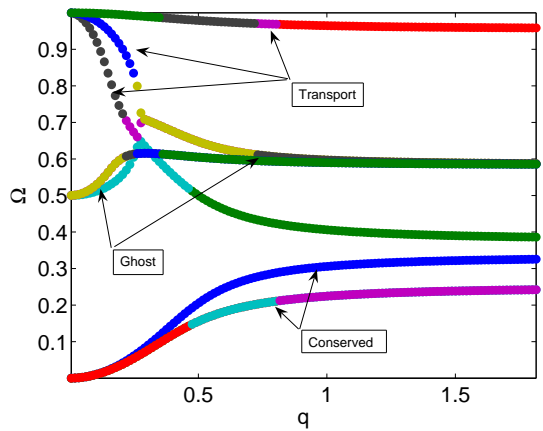}
\caption{Dispersion relation of the $D2Q9$ matrix LBE for $\tau = 1$, $\sigma = 1/2$. The ghost modes relax twice
as fast the stress modes. There is a clean seperation of time scales and enhanced hydrodynamic behaviour compared
to the BGK model.}
\end{figure}

This supports our earlier assertion that the structure of the hydrodynamic and non-hydrodynamic modes 
are dual to each other.  It also illuminates the physical behavior of ghost modes, whose
dynamics appears to be characterized by a competition between global decay as characterised by $\sigma$, 
and instabilities driven by negative diffusion as indicated by the negative curvature of the ghost dispersion relations.
Thus the ghost modes decay globally, but driven by the negative diffusion, concentrate around thinner and thinner
regions of space, and can thereby undermine the high frequency high wavenumber stability of the system.
Good hydrodynamic behavior is thus expected whenever global decay proceeds sufficiently
fast to deplete the ghost energy before this energy has time to cascade to high frequencies.
This picture suggests a number of interesting questions for future studies. First, it would be interesting to explore whether the use entropic methods \cite{Succi:2002}  simply accelerates the ghost decay, or rather turns ghosts into stable modes. Second, following \cite{Sbragaglia:2006a}, it would be interesting to study whether the dual-decomposition can help designing the ghost dynamics in such a way as to absorb energy bursts
from the hydrodynamic component, as they occur in a turbulent flow (intermittency). This could be achieved, for instance, by promoting ghost eigenvalues to dynamical fields responding self-consistently to the local dynamics of the turbulent flow, as it is currently done with the transport eigenvalues $\tau$ in the kinetic modeling of fluid turbulence \cite{Chen:2003}. Finally, we note that the long-wavelength dynamics obtained with the present choice of eigenvectors is, by construction, isotropic at order $k^2$ and Galilean invariant. 

\section{Conclusion}
In this paper we have developed the notion of duality between the hydrodynamic and ghost sectors of lattice kinetic equations, as a guiding criteria to resolve the ambiguities which arise in the practical construction of LB models in matrix form. Our main prescription is that the ghost sector should be constructed, in analogy with the hydrodynamic sector, to consist of density-current pairs. This prescription is exactly realised in the $D2Q9$ model, where in addition, the ghost and hydrodynamic sectors can be interchanged by a suitable swapping of weights. For higher order lattice in and in  higher dimensions the kinetic degrees of freedom are more numerous then the hydrodynamic ones thereby ruling out an exact correspondence between the two.  However, the duality prescription still provides an useful ordering of the eigenvectors into a quasi-dual kinetic space. The duality principle presented  in this paper has been used previously in constructing the kinetic space of the fluctuating lattice Boltzmann equation \cite{Adhikari:2005}. It has also been recently used to compare the accuracy of multireflection boundary conditions with both weighted and un-weighted eigenvectors \cite{Chun:2007}.  We hope the duality principle as introduced here will provide an impetus to further developments in the matrix formulation of the lattice Boltzmann method.

\section{Appendix}

For easy reference we present the eigenvectors of the $D2Q9$ and $D3Q19$ models chosen according to the duality prescription with weighted inner product. The table is arranged according to the conserved (C), transport (T) and ghost (G) sectors. 
$$
A^T = \left[
\begin{array}{l|rrrrrrrrr}
A^{\rho}&1&1&1&1&1&1&1&1&1\\
A^{j_x}&0&1&0&-1&0&1&-1&-1&1\\
A^{j_y}&0&0&1&0&-1&1&1&-1&-1\\
A^{Q_{xx}}&-1&2&-1&2&-1&2&2&2&2\\
A^{Q_{xy}}&0&0&0&0&0&1&-1&1&-1\\
A^{Q_{yy}}&-1&-1&2&-1&2&2&2&2&2\\
A^{\rho^{\prime}}&1&-2&-2&-2&-2&4&4&4&4\\
A^{j^{\prime}_x}&0&-2&0&-2&0&4&-4&-4&4\\
A^{j^{\prime}_y}&0&0&-2&0&2&4&4&-4&-4\\
\end{array}
\right]
$$

\begin{widetext}
$$
A^T = \left[
\begin{array}{l|rrrrrrrrrrrrrrrrrrr}
A^{\rho}& 1 &  1 &  1 &  1 &  1 &  1 &  1 & 
         1 &  1 &  1 &   1 &  1 &  1 &  1 & 1 & 1 & 1 & 1 & 1\\
A^{j_x}& 0 &  1 &  -1 &  0 &  0 &  0 &  0 & 
         1 &  1 &  -1 &   -1 &  1 &  1 &  -1 & -1 & 0 & 0 & 0 & 0\\
A^{j_y}& 0 &  0 &  0 &  1 &  -1 &  0 &  0 & 
         1 &  -1 &  1 &   -1 &  0 &  0 &  0 & 0 & 1 & 1 & -1 & -1\\
A^{j_z} & 0 &  0 &  0 &  0 &  0 &  1 &  -1 & 
         0 &  0 &  0 &   0 &  1 &  -1 &  1 & -1 & 1 & -1 & 1 & -1\\
A^{Q_{xx}}& -1 &  2 &  2 &  -1&  -1 &  -1 &  -1 & 
         2 &  2 &  2 &   2 &  2 &  2 &  2 & 2 & -1 & -1 & -1 & -1\\
A^{Q_{yy}}& -1 &  -1 &  -1 &  2&  2 &  -1 &  -1 & 
         2 &  2 &  2 &   2 &  -1 &  -1 &  -1 & -1 & 2 & 2 & 2 & 2\\
A^{Q_{zz}}& -1 &  -1 &  -1 &  -1&  -1 &  2 &  2 & 
         -1 &  -1 &  -1 &   -1 &  2 &  2 & 2 & 2 & 2 & 2 & 2 & 2\\
A^{Q_{xy}}& 0 &  0 &  0 &  0&  0 &  0 &  0 & 
          1 &  -1 &  -1 &    1 &  0 &  0 & 0 & 0 & 0 & 0 & 0 & 0\\
A^{Q_{yz}}& 0 &  0 &  0 &  0&  0 &  0 &  0 & 
          0 &   0 &   0 &   0 &  0 & 0 & 0 & 0 & 1 & -1 & -1 & 1\\
A^{Q_{zx}}& 0 &  0 &  0 &  0&  0 &  0 &  0 & 
          0 &   0 &   0 &   0 &  1 & -1 & -1 & 1 & 0 & 0 & 0 & 0\\
A^{\rho^{\prime}} & 0 &  1 &  1 &  1 &  1 &  -2 &  -2 & 
         -2 &  -2 &  -2 &  -2 &  1 &  1 & 1 & 1 & 1 & 1 & 1 & 1\\
A^{j^{\prime}_x}& 0 &  1 &  -1 &  0&  0 &  0 &  0 & 
         -2 &  -2 &  2 &  2 &  1 &  1 & -1 & -1 & 0 & 0 & 0 & 0\\
A^{j^{\prime}_y}& 0 &  0 &  0 &  1&  -1 &  0 &  0 & 
         -2 &  2 &  -2 &  2 &  0 &  0 & 0 & 0 & 1 & 1 & -1 & -1\\
A^{j^{\prime}_z}& 0 &  0 &  0 &  0&  0 &  -2 &  2 & 
         0 &  0 &  0 &  0 &  1 &  -1 & 1 & -1 & 1 & -1 & 1 & -1\\
A^{\rho^{\prime\prime}}& 0 &  1 &  1 &  -1&  -1 &  0 &  0 & 
         0 &  0 &  0 &  0 &  -1 &  -1 & -1 & -1 & 1 & 1 & 1 & 1\\
A^{j^{\prime\prime}_x}& 0 &  1 &  -1 &  0&  0 &  0 &  0 & 
         0 &  0 &  0 &  0 &  -1 &  -1 & 1 & 1 & 0 & 0 & 0 & 0\\
A^{j^{\prime\prime}_y}& 0 &  0 &  0 & -1&   1 &  0 &  0 & 
         0 &  0 &  0 &  0 &   0 &  0 & 0 & 0 & 1 &  1 & -1 & -1\\
A^{j^{\prime\prime}_z}& 0 &  0 &  0 &  0&  0 &  0 &  0 & 
         0 &  0 &  0 &  0 &  -1 &  1 & -1 & 1 & 1 & -1 & 1 & -1\\
A^{\rho^{\prime\prime\prime}}& 1 &  -2 &  -2 &  -2&  -2 &  -2 &  -2 & 
         1 &  1 &  1 &  1 &  1 &  1 & 1 & 1 & 1 & 1 & 1 & 1\\
\end{array}
\right]
$$
\end{widetext}

\begin{acknowledgements} SS wishes to acknowledge the Physics Department of the University of  Edinburgh
for financial support and kind hospitality. Both authors wish to thank Prof M. E. Cates for valuable discussions and
a critical reading of the manuscript. RA was funded in part by EPSRC GR/S10377.
\end{acknowledgements}
\bibliography{ghost.bib}

\end{document}